# Microwave Near-Field Imaging of Electric Fields in a Superconducting Microstrip Resonator

Ashfaq S. Thanawalla, S. K. Dutta, C. P. Vlahacos, D. E. Steinhauer,
B. J. Feenstra, Steven M. Anlage, and F. C. Wellstood
*Center for Superconductivity Research, Department of Physics, University of Maryland, College Park, Maryland 20742-4111*
and
Robert B. Hammond
*Superconductor Technology Incorporated, 460 Ward Drive, Suite F, Santa Barbara, California 93111-2310*

## Abstract

We describe the use of a cryogenic near-field scanning microwave microscope to image microwave electric fields from superconducting and normal-metal microstrip resonators. The microscope employs an open-ended coaxial probe and operates from 77 to 300 K in the 0.01-20 GHz frequency range with a spatial resolution of about 200 μm. We describe the operation of the system and present microwave images of Cu and $Tl_2Ba_2CaCu_2O_8$ microstrip resonators, showing standing wave patterns at the fundamental and second harmonic frequencies.

The demonstration of low microwave loss in high-$T_c$ superconductors at 77 K has opened up the possibility for commercial applications which require performance beyond that achievable with normal-metal devices. Unfortunately, present-day high-$T_c$ microwave circuits have shown performance-limiting problems including non-linearity, intermodulation, and power-dependent characteristics.[1] One important class of nonlinearities arises from vortex motion at weak links and grain boundaries[2] and is exacerbated by large current densities found near patterned edges.[1] Most techniques used to characterize these problems involve global measurements which determine the response of the device as a whole.[3-4] To obtain a complete microscopic picture of nonlinearity in superconducting rf devices, spatially resolved probes of microwave current distribution,[5] electric and magnetic field,[6] loss,[7,8] and microstructure are needed. Conventional microwave electric field imaging techniques include using modulated perturbations,[9] electro-optic techniques,[10] scanning force potentiometry,[11] and transmission line probes[12] which locally extract the signal of interest. In this Letter, we describe the imaging of microwave electric fields in superconducting devices using a near-field scanning microwave microscope that operates from 77 to 300 K. We apply this technique to a prototype device, the single-pole microstrip resonant filter, and measure the response of both a normal metal and a superconducting version.

A schematic illustration of our experimental arrangement is shown in Fig. 1. Microwave power is supplied by an HP 83620B signal generator with a frequency range from 0.01 to 20.00 GHz. The power is amplified by an HP 8349B amplifier to about 100 mW and fed to the microstrip resonator through an isolator and a capacitive coupler. We use an open-ended coaxial probe to sense the electric fields from the microstrip. The probe is connected through an isolator to a matched 50 Ω diode which rectifies the rf signal to produce a dc output proportional to power. This dc output is amplified, filtered, and recorded by a computer.

The outer conductor of the coaxial probe shields the inner conductor from lateral components of the field, allowing us to measure mainly the normal component of electric field $E_z$. The field $E_z$ from the microstrip induces a net charge $Q_I$ on the center conductor of the probe.[13] According to Gauss's law, $Q_I = \varepsilon_o E_z A$ where $A \approx 0.03\ mm^2$ is the area of the face of the center conductor. The spatial resolution is limited by the diameter of the center conductor of the probe (200 μm) or the height of the probe above the device, whichever is greater.[14] We can find the absolute magnitude of $E_z$ by relating $Q_I$ to the power P detected at the diode. The total current induced on the probe face by the electric field is $I = dQ_I/dt = i\omega\varepsilon_o E_z A$, where $\omega$ is the frequency. For a lossless transmission line with a characteristic impedance $Z_0 = 50\ \Omega$, the power detected by the matched diode is $P = I^2 Z_0$. Therefore, one finds $E_z = [P/(\omega^2\varepsilon_o^2 A^2 Z_o)]^{1/2}$.

The cryogenic part of the setup (see dashed box in Fig.1) is enclosed in a vacuum can which is inserted in a liquid-nitrogen cooled Cryofab Model CSM 6042 Cryostat. By heating the microstrip, images can be obtained at temperatures from 77 to 300 K. To take an image, the microscope is cooled, the probe height is set, and the diode output is recorded as the microstrip is moved in the x and y directions using a computer-controlled cryogenic two-axis translation stage.

To investigate the capabilities of our system, we imaged a test copper microstrip resonator. The resonator consists of a 350 μm thick circuit board (FR-4 fiberglass) which is coated on both sides with copper. One side serves as a ground plane, while the other is patterned into a microstrip which is 1.1 mm wide and L = 7.6 mm long. The resonator is mounted in a Cu package and microwave signals can be capacitively coupled in and out of the resonator at one end (see Fig. 2a inset).

To determine the resonant frequency of the microstrip, we positioned the probe at one of the open ends (point N in Fig. 1) at a height of 1 mm. We then applied 100 mW of incident power to the capacitive coupler and recorded the diode voltage as a function of frequency. Figure 2a shows the rf power at the diode for temperatures of 77 and 300 K. Each curve shows a single large resonance, some small secondary resonances, and noise; the small resonances are due to reflections off components and connectors within the system. We estimate the loaded quality factor Q of the resonator by fitting the overall frequency response with a single Lorentzian lineshape. We find that the Q of the microstrip increases substantially as the temperature is lowered, mainly due to the improved conductivity of the copper; Q = 40 at 300 K, while Q = 92 at 77 K. From independent broadband transmission measurements at 300 K we determined the Q to be within 10% of the value given above.

At room temperature, the Cu resonator shows a peak in the detected power at about 9.64 GHz, while at 77 K the peak has shifted to 9.95 GHz. Part of the increase in the resonant frequency upon lowering the temperature can be attributed to the decreased inductance from the reduced skin depth (~10 MHz) and thermal contraction of the Cu microstrip (~20 MHz).[15] From independent capacitance measurements, we found that the 310 MHz total shift is mainly due to the dielectric constant of the fiberglass substrate decreasing by 10% upon cooling.

The 9.5 - 10 GHz resonance in Fig. 2a corresponds to the fundamental standing wave mode for the microstrip. This is demonstrated in Fig. 3a, which shows an image of the magnitude of $E_z$ above the Cu microstrip resonator at 77 K. During the scan, the frequency of the source was fixed at the resonant frequency (9.95 GHz) and the probe was held about 1 mm above the microstrip. The image shows a maximum signal at the two open ends of the microstrip conductor (corresponding to the voltage antinodes), and a minimum in the middle (corresponding to the voltage node), as expected.

We next imaged a $Tl_2Ba_2CaCu_2O_8$ (TBCCO) microstrip resonator.[16] The resonator has a 650 nm thick film of TBCCO deposited on both sides of a 420 μm thick, 4.7 mm x 9.9 mm MgO substrate. One side serves as a ground plane, while the other side is patterned into a microstrip with a width of 150 μm and a length L = 7.1 mm. We mounted the resonator in a Cu package and capacitively coupled rf power to one end. Figure 2b shows a frequency scan taken at 77 K (at point N in Fig. 1), with the probe at a height of about 250 μm above the microstrip. Note the much smaller frequency range, as compared to Fig. 2a, indicating the substantially higher Q of the superconducting resonator.

At 77 K we find a resonance at about 8.205 GHz. The resonance shifts to lower frequencies (see Fig. 2b inset) and diminishes in magnitude as the temperature is raised toward $T_c$ ~ 103 K. No resonance was found above $T_c$. The decrease in resonant frequency at higher temperatures is due to the increase in penetration depth $\lambda$ of the superconductor. The temperature dependence of the resonant frequency $f_0$ was fit using a BCS s-wave temperature dependence[17] for $\lambda$ (T) (solid line in Fig. 2b inset). The fit yields $T_c$ =

103.2 K and λ(0) = 0.24 μm. The $T_c$ is in good agreement with separate ac-susceptibility measurements (101 - 103 K), and λ(0) is in good agreement with an independent determination by Willemsen *et al.* (λ(0) = 0.26 μm)[18] in similar TBCCO thin films. The loaded Q goes from about 1200 at 77 K to about 480 at 100 K, with a temperature dependence very similar to the one observed for $f_0$. The measured Q at 77 K allows us to put an upper bound on the surface resistance of TBCCO: $R_s$ • 11.3 mΩ at 8.205 GHz. However, the Q is about a factor of 2 lower than the value measured using a network analyzer in the absence of the probe, suggesting that coupling to the probe contributes significant loss.

The lower image in Fig. 3b shows $|E_z|$ above the TBCCO microstrip resonator, measured at 77 K. The frequency of the source is fixed at the resonant frequency (8.205 GHz) and the probe is 180 μm above the microstrip. As with the Cu resonator, we observe a maximum electric field at the two open ends, as expected for the fundamental mode. However, the field strength is somewhat different at the two antinodes. This is probably due to the presence of the input capacitive coupling pin on the resonator's left hand side, which perturbs the strip's simple geometry and also affects the perturbation of the resonance frequency by the probe. In addition, note that $E_z$ is much larger than in Fig. 3a, due to the higher Q of the TBCCO microstrip and the smaller probe-sample separation.

The issue of probe perturbation is relevant to all quantitative electric field measurement techniques[9-12] and is a topic of ongoing research. In order to evaluate the extent to which perturbations of the resonant frequency affect our images, we also imaged the electric field in a frequency-following mode. In this mode, the frequency of the source is kept on the resonance of the microstrip as the probe is scanned across the resonator.[19] Clearly this technique will not work at a node, where no signal is present. The upper image in Fig. 3b shows the result for such a scan, measured at the right antinode of the fundamental mode for the TBCCO microstrip at a height of 230 μm. In comparison to the fixed-frequency image, $E_z$ shows a different curvature, producing less peaked antinodes. The difference between the two images is due to shifting of the high Q resonance frequency by the probe, which is not accounted for in the fixed frequency image. On the other hand, neither image shows the $\cos^2(\pi x/L)$ dependence seen in Fig. 3a. This is mainly due to the smaller aspect ratio of the TBCCO microstrip, which tends to concentrate the electric field at the ends.

For comparison, Fig. 3c shows an electric field image obtained at 16.2 GHz, where there is also a peak in the frequency response of the microstrip. For this image the probe was held 250 μm above the microstrip. We see a complete standing wave pattern with antinodes at the two open ends and the middle of the microstrip, as expected for the second harmonic resonance. The TBCCO strip appears to be fairly well resolved, consistent with a spatial resolution on the order of the 200 μm probe size and the probe-sample separation. Finer tips and smaller probe-sample separations can be used to achieve higher spatial resolution.

In conclusion, we have demonstrated a simple broadband microwave imaging system which allows us to measure the normal component of electric field near an operating high frequency device at cryogenic temperatures. The system has achieved about 200 μm spatial resolution using only simple microwave instrumentation combined with a cryogenic scanner, establishing the basis for a microscopic examination of superconducting microwave devices.

We wish to gratefully acknowledge R. Newrock for fabrication of the cryogenic scanner and Lucia V. Mercaldo for the penetration depth analysis. This work has been supported by the National Science Foundation NSF grant No. ECS-9632811, NSF-MRSEC grant No. DMR-9632521, and by the Maryland Center for Superconductivity Research.

# Figure Captions

Fig. 1. Schematic illustration of the cryogenic microwave microscope. The dashed box indicates the cryogenic section.

Fig. 2. (a) rf Power at the diode detector vs. frequency for the Cu microstrip at 77 and 300 K. Inset shows schematic of the microstrip and Cu package. (b) rf Power at the diode vs. frequency for TBCCO microstrip at 77 and 100 K. The corresponding magnitude of the normal component of electric field at the probe is indicated on the right axes. Inset shows temperature dependence of the resonant frequency $f_0$ along with a fit to BCS-theory (solid line).

Fig. 3. (a) Magnitude of the normal component of electric field $E_z$ above a Cu microstrip excited at its fundamental frequency (f = 9.95 GHz). (b) $E_z$ above a TBCCO microstrip excited at its fundamental frequency (f = 8.205 GHz, lower image). For comparison, the upper image shows $E_z$ measured using a frequency following technique. (c) $E_z$ above a TBCCO microstrip excited at its second harmonic frequency (f = 16.2 GHz). The contour lines are labeled with field strength in V/mm. In all cases, power is fed into the microstrip on the left hand side and the temperature is 77 K.

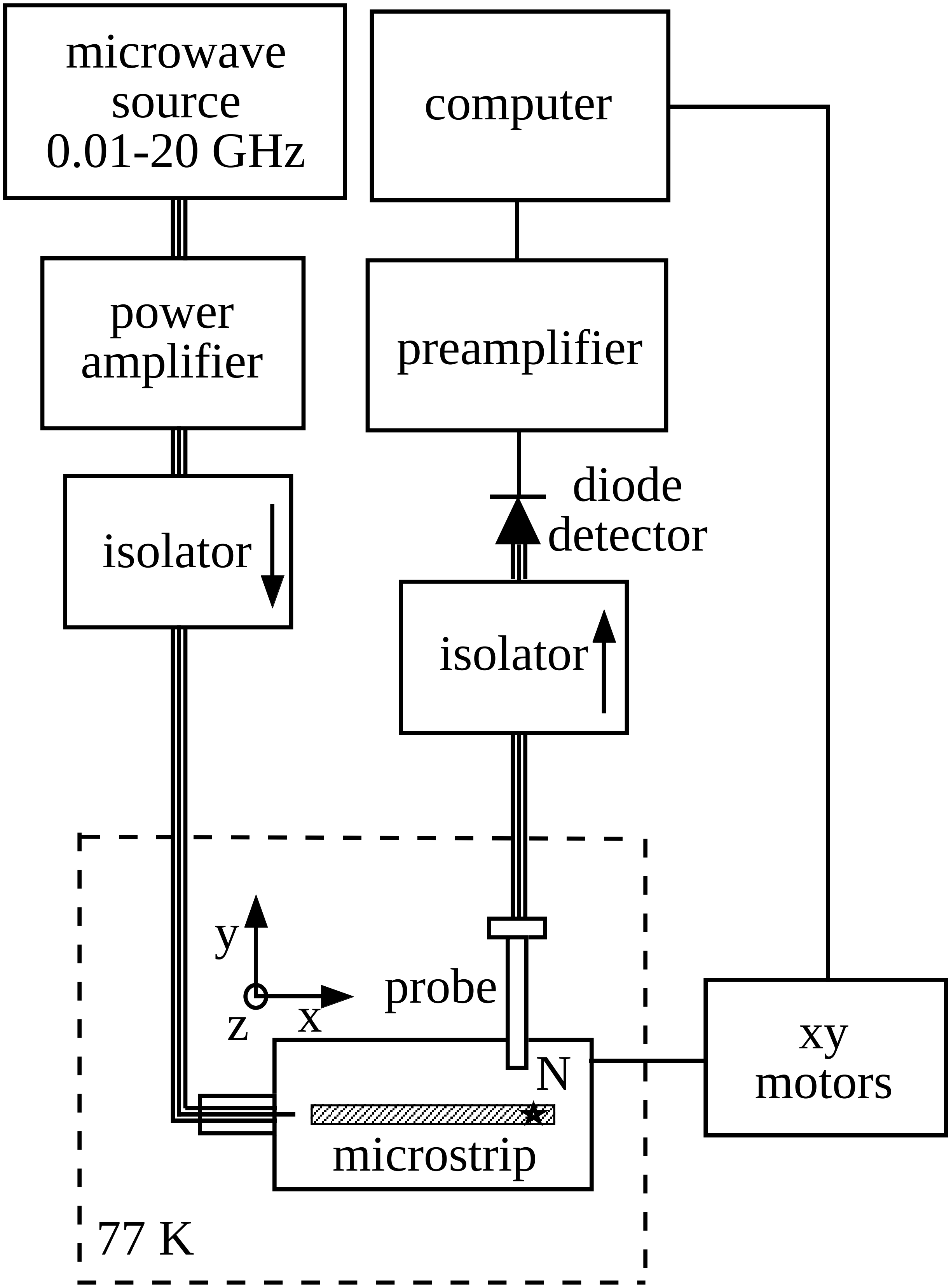

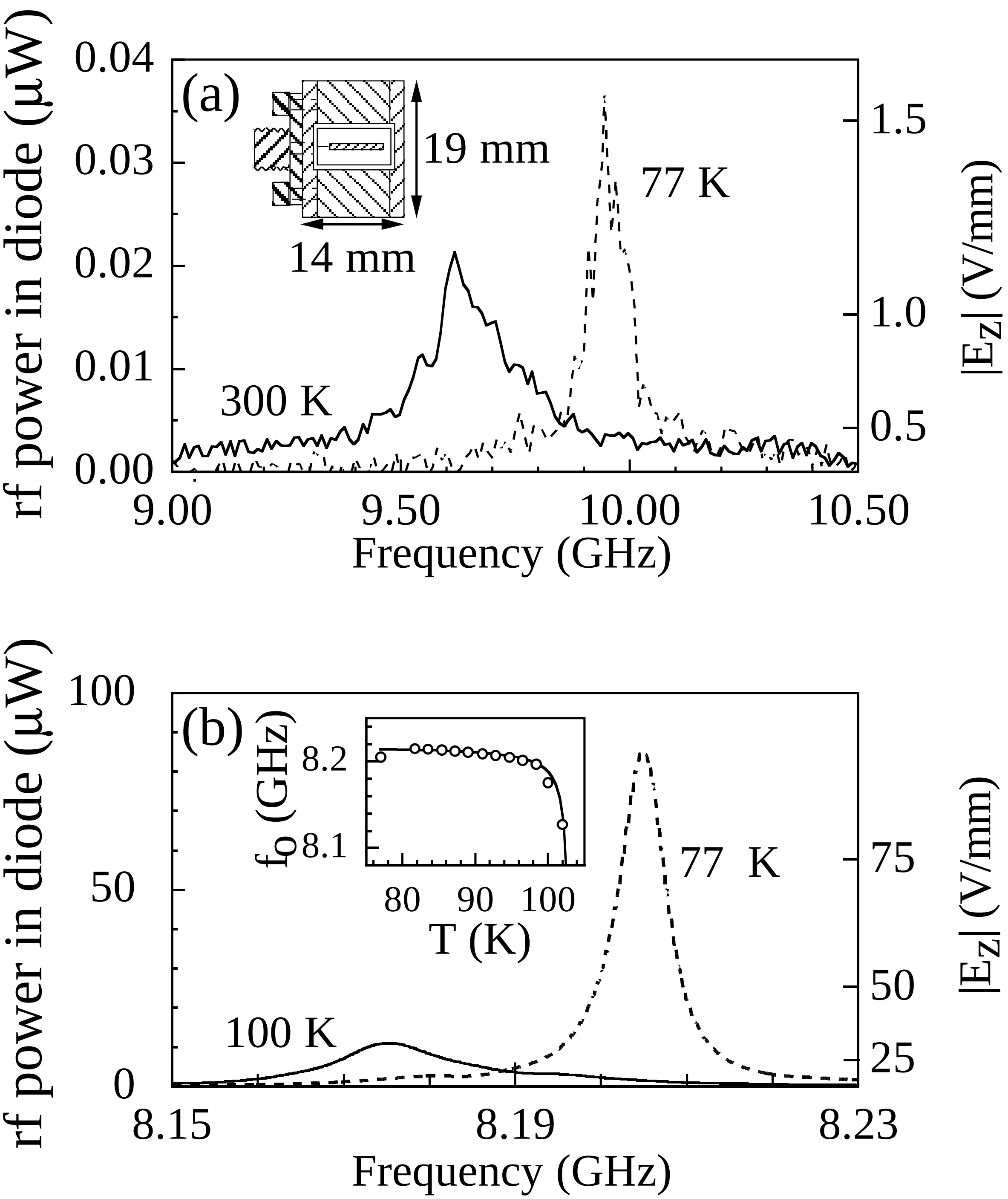

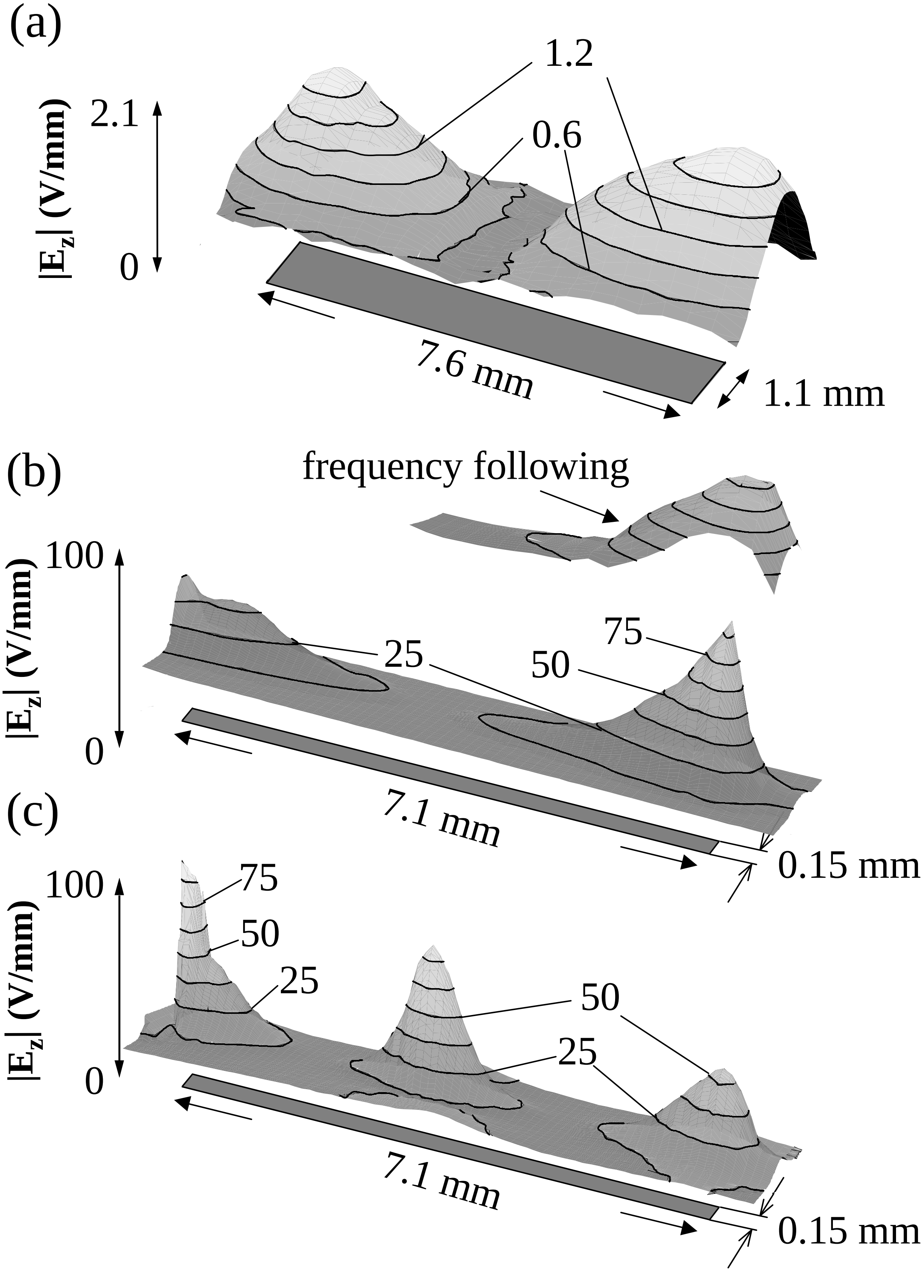

Fig 3, Thanawalla *et al*.